\documentclass{raa}            
\usepackage{graphicx,times}
\usepackage{natbib}

\providecommand{\bm}[1]{\mbox{\boldmath$#1$\unboldmath}}
\def\kms{$\rm{km~s}^{-1}$}
\def\etal{et al.}
\def\hmpc{$h^{-1}$~Mpc}

\begin{document}

  \title{Comprehensive Spectral Analysis of Cyg  X-1 using RXTE Data
}

 \volnopage{ {\bf 2010} Vol.\ {\bf X} No. {\bf XX}, 000--000}
   \setcounter{page}{1}

   \author{Rizwan Shahid
      \inst{1}
   \and R. Misra
      \inst{2}
   \and S. N. A. Jaaffrey
      \inst{1}
   }

   \institute{Dept. of Physics, M. L. Sukhadia University, Udaipur-313001, 
             India; \\
        \and
             Inter University Centre for Astronomy and Astrophysics,
              Pune-411007, India; {\it rmisra@iucaa.ernet.in}\\
\vs \no
   {\small Received [year] [month] [day]; accepted [year] [month] [day] }
}

\abstract{We analyse a large number ($> 500$) pointed RXTE observations of Cyg X-1 and model the spectrum
of each one. A subset of the observations for which there is simultaneous reliable measure of the
hardness ratio by the All Sky Monitor, shows that the sample covers nearly all the 
spectral shapes of Cyg X-1.  The relative strength, width of the Iron line and
the reflection parameter are in general correlated with the 
high energy photon spectral index $\Gamma$.
This is broadly consistent with a geometry where for the hard state (low $\Gamma \sim 1.7$) there
is a hot inner Comptonizing region surrounded by a truncated cold disk. The inner edge of the
disk moves inwards as the source becomes softer till finally in the soft state (high $\Gamma > 2.2$)
the disk fills the inner region and active regions above the disk produce the 
Comptonized component. However, the reflection parameter shows non-monotonic behaviour near the
transition region ($\Gamma \sim 2$), suggestive of a more complex geometry or physical 
state of the reflector. Additionally, the inner disk temperature, during the hard state, 
 is on the average higher than in the soft one , albeit with large scatter.  These inconsistencies could be due
to limitations in the data and the empirical model used to fit them.  The flux of each spectral component is well
correlated with $\Gamma$ which shows that unlike some other black hole systems, Cyg X-1 does
not show any hysteresis behaviour. In the soft state, the flux of the Comptonized component is
always similar to the disk one, which confirms that the ultra-soft state (seen in other brighter
black hole systems) is not exhibited by Cyg X-1.  The rapid variation of the Compton
Amplification factor with $\Gamma$, naturally explains the absence of spectra with $\Gamma < 1.6$,
despite a large number having $\Gamma \sim 1.65$. 
\keywords{accretion, accretion disks: X-rays: binaries}
}

   \authorrunning{Shahid, Misra \& Jaaffrey }            
   \titlerunning{Comprehensive Spectral Analysis of Cyg  X-1 using RXTE Data }  
   \maketitle

\section{Introduction}
\label{sect:intro}
Over the last forty years, several generations of X-ray satellites have observed Cyg X-1, to
make it one of the best studied black hole systems. The source is known to undergo spectral state
transitions between a hard/low state and a soft/high state. Historically, most of the times the source
is found in the hard state which can be characterised approximately as a hard power-law emission 
with a high energy cutoff at $\sim 100$ keV. The basic model invoked to explain the hard state
is that of a truncated cold standard accretion disk with an hot inner region \citep{Sha76}. In this
geometry photons from the outer truncated disk impinge into the hot inner region and produce the
observed Comptonization spectrum. This
generic model has been refined by the formulation of theoretically more consistent and stable hot
inner disks \citep[e.g.][]{Nar94,Esi98} and by more detailed spectral modelling which include
the effect of reflection and possible non-thermal emission \citep[e.g.][]{Gie97}. An alternate
interpretation of the hard state is that there is an extended transition region where
the temperature increases rapidly and the emergent spectrum is the sum of the local spectra
of each radii \citep{Mis97,Mis00}.

In contrast,
the soft state is dominated by a thermal emission and a steep hard X-ray emission extending
to at least several hundred keV. In analogy with the solar corona, the basic model for the
soft state is that of a cold accretion disk extending to the last stable orbit with a
hot corona on top \citep{Lia77}. Theoretically more consistent modelling comprising of 
active regions or blobs on top of the cold disk have been formulated \citep[e.g.][]{Haa94} and
detailed spectral modelling of this state including reflection and non-thermal emission
have been undertaken \citep{Gie99}. \cite{Don07} present a review of the present understanding
of the accretion process in such systems.

Cyg X-1 shows these two spectral states and transitions between them which are sometimes
referred to as an  intermediate state \citep[e.g.][]{Mal06}. Other
black holes systems,  especially GRS 1915+105
show a myriad of spectral states whose classification also depends on the temporal
property of the source \citep[e.g.][]{Bel97,Dun10,Zdz04,Rem06}. Conspicuous with its 
absence for Cyg X-1 seems to be  the very high or ultra-soft state where the spectrum is 
dominated by the thermal emission and the power-law emission is weak. Over all,
the spectral behaviour of Cyg X-1 is simpler than other black hole systems and 
hence its systematic analysis may pave the way for a better understanding of
black hole systems in general.

For more than a decade the {\it Rossi X-ray Timing Experiment} (RXTE), with its
Proportional Counter Array (PCA) and High Energy X-ray Timing Experiment (HEXTE), has provided
unprecedented coverage of Cyg X-1. The high time resolution and large effective area of the PCA
has provided a wealth of information regarding the rapid variability of black hole systems.
For Cyg X-1, the shape of the power spectra and the discovery of frequency dependent time lags between different energy bands
\citep{Now99} have indicated that the variability is due to propagation of waves in the
disk \citep{Lyu97,Now99,Tit07,Mis08,Mis00,Kot01}. Detailed power spectral analysis of Cyg X-1 
using a number of RXTE observations have been undertaken \citep{Pot03,Axe05}.

Detailed photon spectral study of specific observations of Cyg X-1 by RXTE have been undertaken
in the hard, soft states and also in the transition phase
\citep[e.g.][]{Gie97,Gie99,Mal06,Now11}. The large number of observations of the
source by RXTE allows for a more comprehensive study of its spectral properties and
correlations between them. \cite{Ibr05} have undertaken a detailed spectral study of
42 {\it Ginga}-RXTE and RXTE-OSSE observations, but these observations were limited
to the hard state. Recently, in a brief report, \cite{Gie10} have made a detailed
spectral analysis and the timing properties of 33 representative RXTE observations 
covering the different spectral shapes of Cyg X-1.

\cite{Wil06} have analysed the spectra of $\sim 200$ pointed RXTE observations of Cyg X-1,
and have fitted them with an empirical broken power-law and thermal Comptonization models,
with reflection and Iron line emission. They caution that since the empirical broken power-law
model is a reasonable fit to the data, more complicated models could be case of "overparametezing".
In particular, they state that the broad Iron line feature could be an artifact caused by fitting
an incorrect simple thermal model to the soft excess component. This indeed is true given the
energy resolution of the RXTE instruments and the systematic uncertainties in their
response. However, as discussed above, in the standard paradigm, the standard accretion disks 
in black hole systems extends to the last stable orbit in the soft state and a broad Iron line
in expected from such a system. Moreover, although \cite{Wil06} have analysed $\sim 200$ 
observations, these may not cover the entire range of spectral shapes exhibited by Cyg X-1.   

In this work, we analyse 504 pointed RXTE observations of Cyg X-1 and fit the spectra
of each one with a uniform model. We show that these large number of observations represent
nearly all the spectral shapes of Cyg X-1. Our motivation is to verify the standard paradigms
of the geometry of the hard and soft states, by studying the correlation between different
spectral components and to bring out any discrepancies or complexities in the spectral evolution.

\begin{figure}
\centering
\includegraphics[width=\textwidth, angle=0]{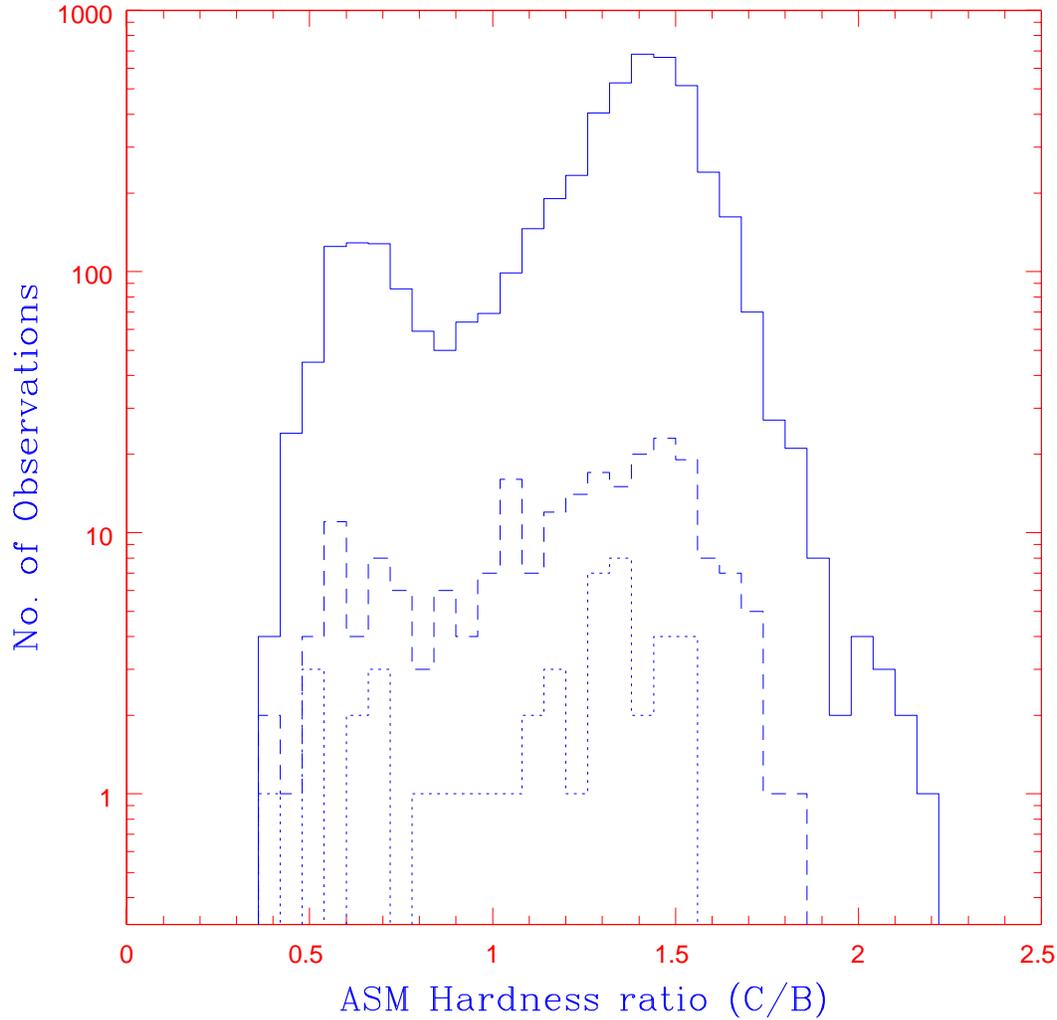}
\caption{The distributions of the ASM hardness
ratio, $H  =C/B$ between the C   (5 - 12.2 keV) and B (3 - 5 keV) energy bands. The solid
line represents all the one-day average detections for which $H$ has 
 a 1-sigma error less than 10\%. The dashed line represents 221 of the 501 pointed observations, analysed
in this work,  for
which there is simultaneous ASM dwell data and for which the average hardness in the ASM
band have errors less than 10\%. These observations scan nearly the entire range
of hardness detected by the ASM. The dotted line represents 
the 49 of the 200 pointed observations of \cite{Wil06} analysis 
 for which there is simultaneous ASM dwell data. }
\label{asmhist}
\end{figure}

\section{The RXTE sample}
Over the last 14 years, the {\it All-Sky Monitor} (ASM) of  RXTE has provided
extensive coverage of Cyg X-1. We obtained the ASM light-curves from 
http://xte.mit.edu/ASM\_lc.html, using the standard filtering criteria mentioned
there. One-day average light-curves were obtained and the hardness ratio, $H$,
between the C  (5 - 12.2 keV) and B (3 - 5 keV) were computed. For 4781 
daily averages, the
hardness ratios have  a 1-sigma error less than 10\% and the histogram of these
data are shown in Figure \ref{asmhist} (solid line). There are two clear
peaks in the distribution corresponding to the hard ($H \sim 1.5$)
and soft  ($H \sim 0.6$) states. 

In this work, we have analysed the spectra of 504 pointed observations of RXTE.
 221 of these observations have simultaneous ASM dwell data, whose average hardness
ratio could be estimated with an uncertainty less than 10\%. The distribution
of their hardness ratio is plotted in Figure \ref{asmhist} (dashed line). The figure
shows that the range  $0.35 <H < 1.85$ covered by these pointed observations are
nearly the entire range observed by ASM, except for some rare observations 
with $H > 1.8$. There are no simultaneous pointed observations during the time
when the ASM detected   $H > 1.8$. Thus the sample of  pointed observations 
analysed in this work covers nearly the entire range of spectral variability of
Cyg X-1. As a comparison, the dotted line in Figure \ref{asmhist} represents
the 49 observations of the sample analysed by \cite{Wil06} which have simultaneous
ASM data.

The pointed observations which have simultaneous ASM detections can be used to
calibrate the ASM count rate and to understand any time-dependent systematics that
the instrument may have. This in turn, will allow for the study of the long term
(days to months) variability of the source in conjunction with the spectral properties.
 We defer that exercise to a latter work, and here we concentrate instead on the PCA/HEXTE spectral analysis. 

\begin{figure}
\centering
\includegraphics[height=80mm, angle=0]{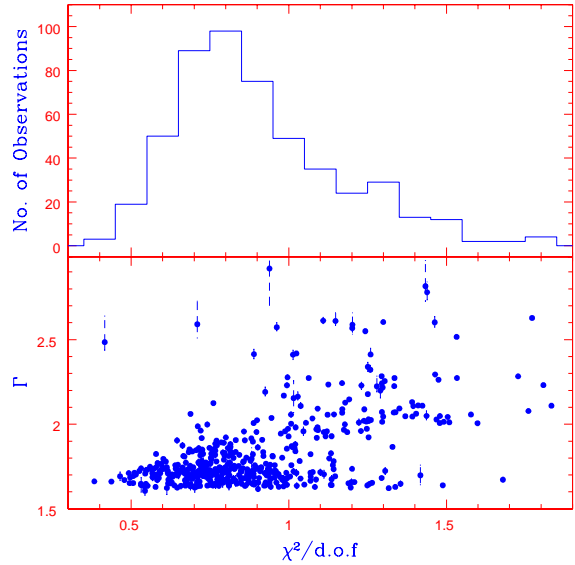}
\caption{Top Panel: The distribution of observations as a function of reduced $\chi^2_{red} = 
\chi^2/d.o.f$. The peak
of distribution is at $\sim 0.8$ and for all observations $\chi^2_{red} < 1.9$. Bottom panel:
The variation of the photon spectral index $\Gamma$ with $\chi^2_{red}$.  }
\label{histchi}
\end{figure}

\begin{figure}
\centering
\includegraphics[height=80mm, angle=0]{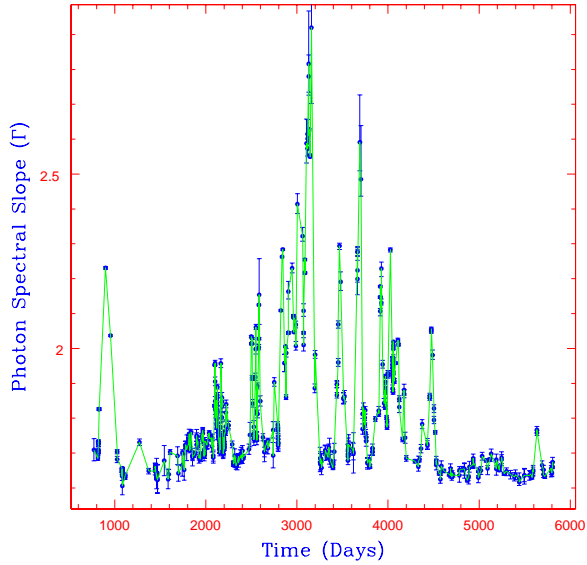}
\caption{ The variation of Photon spectral index with time for the pointed PCA observations
analysed in this work. }
\label{time_alp}
\end{figure}

\section{Spectral Analysis}

For  504 RXTE pointed observations of Cyg X-1, we use the spectral and
response files for the PCA and HEXTE data, 
that have been generated for the standard products using a   
general script\footnote{http://heasarc.gsfc.nasa.gov/docs/xte/recipes/stdprod\_guide.html}.
While the general script may not be optimised for a single observation, it is 
appropriate for this work, where the sample properties of a large number of observations
are being studied. For the PCA data we fit in the energy range $3$-$20$ keV, while
for HEXTE the chosen range was $20$-$80$ keV. A systematic uncertainty 
of $0.01$ was included in all fits \citep{Jah06}.

All the observations were fitted by a generic model consisting of a multi-colour disk black
body and a hot thermal plasma which Comptonizes the disk photons. 
The model includes reflection and an Iron line.  In terms of XSPEC
routines the model is described as  
{\it wabs(diskbb + reflect(nthcomp) + Gaussian)}. Since the
PCA energy band is $> 3$ keV, the column density of the absorption component
(wabs) cannot be constrained and hence was fixed at a negligible value of $10^{21}$ cm$^{-2}$. The
disk black body emission ({\it diskbb}) is parametrised by the inner disk temperature,
$kT_{in}$.  In the thermal Comptonization model, {\it nthcomp} \citep{Zdz96, Zyc99}, the input photon spectrum
is taken to be the disk black body shape and its temperature is tied to $kT_{in}$. The
thermal Comptonized spectrum is parametrised by the electron temperature $kT_e$ and
the high energy spectral index, $\Gamma$. Since HEXTE is not sensitive enough to detect the
high energy turnover in the spectrum, the temperature was fixed at $kT_e = 100$ keV.
We have checked that our results are insensitive as long as  $kT_e > 50$ keV. The Iron
line was represented by a variable width Gaussian with centroid energy fixed at $6.4$ keV.
 Finally, a convolution model ({\it reflect}) produces the reflected component
of the incident X-ray photons from the accretion disk. Apart from the three normalisation
factors for the additive models, the parameters of this generic model are the inner
disk temperature $kT_{in}$, the high energy photon spectral index $\Gamma$, the reflection
fraction $R$ and the width of the Gaussian line, $\sigma$. The model was fit
to all the observations and errors ($\Delta \chi^2 = 2.7$) on these parameters were computed.
The presence of the convolution model for reflection and the large number of observations
made the spectral fitting process, a computer intensive and extremely time consuming one.

Although physically motivated the generic model is basic and does not incorporate
several important effects.  For example, the  intrinsic absorption in the source could
be as high as a few $\times 10^{21}$ cm$^{-2}$ \citep[e.g.][]{Fen02}, while we do not
consider the effect here. Moreover,  there could be a component of non-thermal
electrons in the hot plasma, especially in the soft state of the source. The reflected
component is assumed to be from neutral material while in general it maybe ionised.
The Iron line modelled as a Gaussian is an approximation to more realistic models
which take into account the skewed nature of the line profile due to gravitational effects.
However, as shown below, given the spectral resolution of the PCA and HEXTE, more sophisticated models
like the above are poorly constrained and hence not warranted in this analysis. Nevertheless, 
as always, the spectral resolution of the detectors and the simplifying assumptions of the
spectral models, should be taken into account, when one interprets the results of the analysis.

 The top panel of Figure \ref{histchi} shows the distribution of the observations as a function of the
reduced $\chi^2_{red} = \chi^2/d.o.f$. The peak
of distribution is at $\sim 0.8$ and for all observations $\chi^2_{red} < 1.9$ The
degrees of freedom ($d.o.f \sim 80$) is large and this distribution ideally should be a
narrow Gaussian centred at unity. However, the peak of the distribution is at $\sim 0.8$ instead
of $\sim 1$. On the other hand, there are a few observations where $\chi^2_{red}$ is too high. 
If we include more spectral components or parameters to the model, these high values of 
$\chi^2_{red}$ may become acceptable, but the peak of the distribution will
shift still lower than $0.8$ and the analysis would suffer severely from over-modelling. 
Hence, the level  of model sophistication used here may be close to optimal. 
To see if the observations with large values of $\chi^2_{red}$ are biasing the analysis, we plot in 
the bottom panel of
of Figure \ref{histchi}, the best fit $\Gamma$ values for the corresponding $\chi^2_{red}$.
While there is a tendency for spectra with higher $\Gamma$ to have larger $\chi^2_{red}$, the
effect is not severe. We have checked that if the observations with  $\chi^2_{red} > 1.5$ 
are removed from the sample, the results described in the next section are not
affected.

\begin{figure}
\centering
\includegraphics[height=80mm, angle=0]{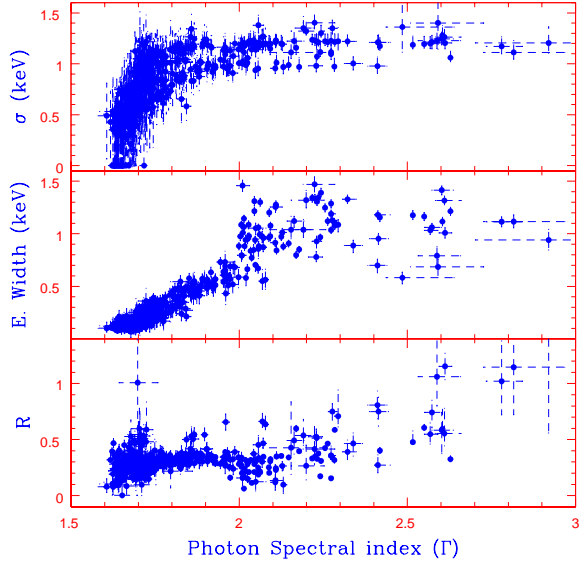}
\caption{ The width  $\sigma$ (top panel), the Equivalent Width (middle panel) of the Iron line
and the reflection parameter $R$  versus
the photons spectral index $\Gamma$. For three observations $R > 2$ and hence are not plotted
in the bottom panel.}
\label{gam_Fe}
\end{figure}

\begin{figure}
\centering
\includegraphics[height=80mm, angle=0]{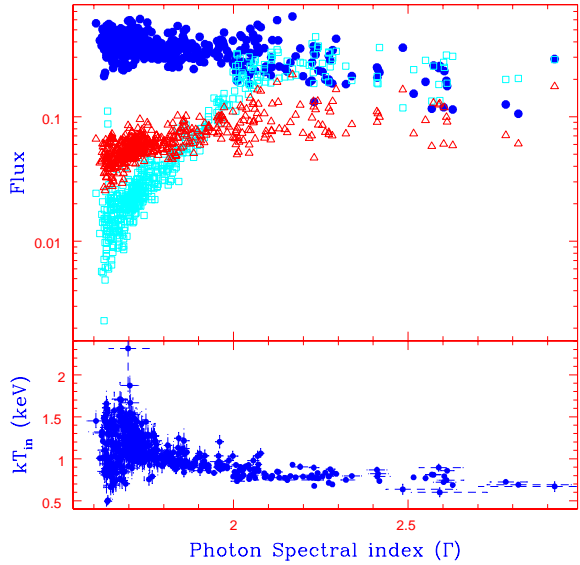}
\caption{Top panel: The bolo-metric flux ($10^{-7}$ ergs/cm$^2$/sec) versus
the photons spectral index $\Gamma$. The circles represent the  flux of the
Comptonization component, $F_c$ while  the squares represent the  flux of the disk black body
component, $F_{DBB}$. The triangles represent the flux of the input photons entering the
Comptonizing region, $F_{in}$. Bottom panel: The inner disk temperature, $kT_{in}$ versus $\Gamma$. 
  }
\label{gam_Flux}
\end{figure}

\begin{figure}
\centering
\includegraphics[height=80mm, angle=0]{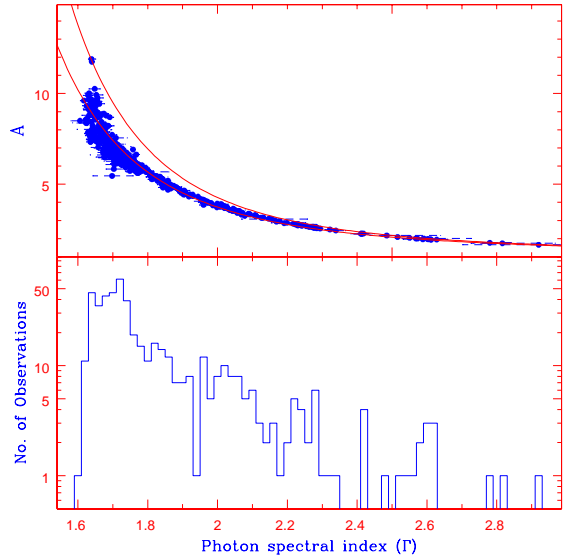}
\caption{Top panel: The Compton amplification factor $A = F_c/F_{in}$ as a function of $\Gamma$. The lines
represent the cases when $kT_{in} = 1.0$ keV (bottom line) and $ = 0.5$ keV (upper line). Bottom Panel: The distribution of observations as a function of $\Gamma$. Note the sharp decline in number for $\Gamma < 1.65$ 
coinciding with the steep increase in the Amplification factor.   }
\label{gamA}
\end{figure}

\section{Results}

The high energy photon index, $\Gamma$ of the thermal Comptonized component was the best
constrained parameter and hence we use it as a basis and study its correlation with other parameters.
Moreover, $\Gamma$ is a good indicator of the spectral state of the system and changes smoothly
from the low to hard states \citep{Gie10}. For the PCA data analysed here we plot $\Gamma$ versus time in
Figure \ref{time_alp}. The data covers a wide time range and samples the various state transitions
as well as the "failed" or partial transitions when $\Gamma < 2$.

The width  $\sigma$, the relative strength  of the Iron line
and the reflection parameter $R$  have been plotted versus $\Gamma$ in Figure \ref{gam_Fe}.
The relative strength of the Iron line (i.e. the Equivalent Width) correlates tightly
with $\Gamma$ for $\Gamma < 2.0$. There is a sharp change of behaviour at $\Gamma \sim 2$
and for larger $\Gamma$ values the Equivalent Width is uncorrelated and has  a larger
dispersion. The width of the line shows a similar behaviour being correlated at low
$\Gamma < 1.8$ and is uncorrelated for larger values. There is a hint of double valued
solutions for the width versus $\Gamma$ but given the errors it is difficult to make 
any concrete statements. If the Iron line emission and the reflection component are from
the same physical component, they should be correlated and indeed the reflection parameter
$R$ is also broadly correlated with $\Gamma$ as expected. However, the dependence is more complex
with $R$ having a non-monotonic behaviour i.e. there seems to be anti-correlation when
$1.8 < \Gamma < 2.1$.

The top panel of Figure \ref{gam_Flux} shows the unabsorbed bolo-metric flux of the 
Comptonization component, $F_c$ (filled circles), disk black body
component, $F_{DBB}$ (open squares) and that of the  input photons entering the
Comptonizing region, $F_{in}$ (open triangles). As expected, the Comptonization component
dominates during the hard state, while in the soft state, its flux is comparable
with that of the disk black body. It is interesting to note that the flux
of the disk black-body photons entering the Comptonization region is smaller
than the observed disk black body flux in the soft state, but is larger in
the hard state. This may indicate changes in the geometry of the system as
discussed in the next section. The variation of the inner disk temperature
$kT_{in}$ is shown in the bottom panel of Figure \ref{gam_Flux}. The
temperature is nearly constant with a hint of an increase in the
hard state.

The top panel of Figure \ref{gamA} shows the Compton Amplification factor $A = F_c/F_{in}$ which 
depends on $\Gamma$, $kT_{in}$ and the hot electron temperature $kT_e$. In this
modelling, $kT_e = 100$ keV is fixed. The solid lines depict the variation of
$A$ for a fixed value of $kT_{in} = 1.0$ keV (upper curve) and $ 0.5$ keV (lower
curve). Variations in $kT_e$ would shift these curves vertically. The estimated
variation of $A$ rises more steeply for smaller $\Gamma$ than for the case
when $kT_{in}$ is a constant. This implies that in the hard state, as the
required amplification increases, the system adjusts by decreasing $\Gamma$ and
$kT_{in}$. The bottom panel of the Figure shows the distribution of observations with
$\Gamma$.  Note that there is sharp decline in the number of sources having
$\Gamma < 1.65$ which is coincident with the steep increase in the Amplification
factor.

These results are broadly consistent with those of \cite{Wil06}. In particular,
for their thermal Comptonization model fitting, the width and strength of
the Iron line increases and saturates with decreasing Compton y-parameter, which
is similar to the behaviour described here as a function of the photon spectral index.
They also find a similar trend for the reflection parameter $R$, however their results
do not show any  non-monotonic behaviour as found in this work which may be due to the larger number of 
observations analysed in this work.
 
\begin{figure}
\centering
\includegraphics[height=80mm, angle=0]{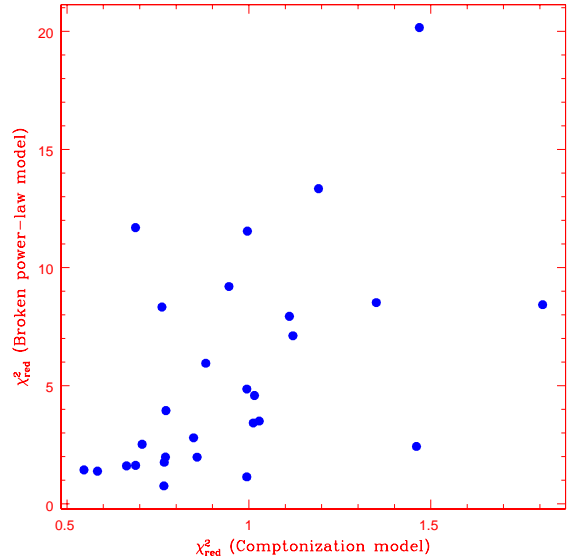}
\caption{Comparison of reduced $\chi^2$ for the Comptonization model and the broken power-law one for 23 Observations. The
reduced $\chi^2$ for broken power-law is larger than the power-law, but the systematic uncertainty  of RXTE response
 does not allow for any concrete statements to be made.    }
\label{chi_comp}
\end{figure}

\begin{figure}
\centering
\includegraphics[height=80mm, angle=0]{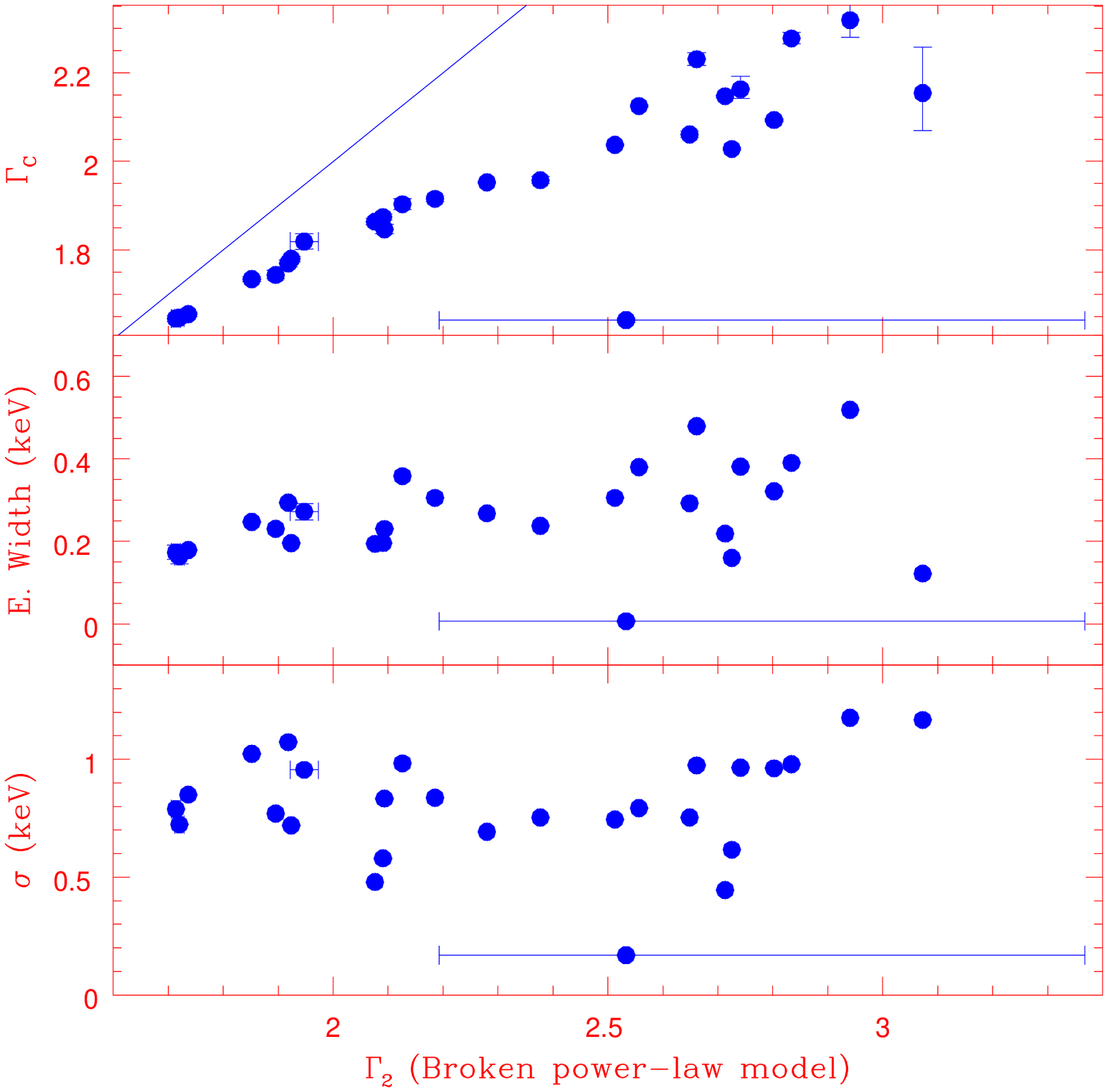}
\caption{Top Panel: The variation of the photon spectra index $\Gamma_c$ from the Comptonization model
used in this work with the high energy spectra index of the broken power-law model. While there is 
the expected correlation, the spectral index of the broken power-law model is systematically higher than
the Comptonization one. Middle and Bottom panel: The variation of the strength (EW) and the width of the
Iron line versus the high energy power-law index for the broken power-law model. There is no clear correlation
as was seen for the Comptonization model.      }
\label{comb_gam}
\end{figure}

To re-emphasis that the results obtained here are subject to the validity of the Comptonization model used,
we have analysed 27 representative data sets with a broken power-law and a Gaussian model. The 23 observations represent
the complete range of spectral index $\Gamma$ obtained from the Comptonization model. Figure \ref{chi_comp} compares the
reduced $\chi^2$ for the broken power-law model and the Comptonization one used here. Clearly, the broken power-law model
gives significantly larger reduced $\chi^2$ than the Comptonization one. However as pointed out by \cite{Wil06}, the systematic
uncertainties of the PCA response does not allow for any concrete statements that broken power-law model can be rejected. 
The top panel of Figure \ref{comb_gam} shows the variation of the high energy photon index of the broken power-law model versus
that of the Comptonization model. There is the expected correlation, but the broken power-law model indices are systematically higher.
The middle and bottom panels of \ref{comb_gam} show the strength and width of the Iron line versus the high energy photon index and
no clear correlation can be seen. Thus, the correlations found in this work, for the Comptonization model, should be considered only
within the framework of the model used. In other words, if the Comptonization model is assumed to be a good representative of the
real physical radiative process, then the data suggests that there are correlation between the Iron line parameters and the
spectral index (Figure \ref{gam_Fe}).

\section{Discussion}

An attractive and more or less standard model for Cyg X-1 is a geometrical one
where for the hard state, there is a truncated disk surrounding a hot Comptonizing
region in contrast to the soft state, where the disk extends to the inner regions
and a hot corona on top Comptonizes its photons. During a state transition the
inner radius of the truncated disk moves inwards, while the inner Comptonizing 
region shrinks. As the cold disk fills the inner regions, a hot corona (or more 
specifically several active regions) arise on top of it and Comptonize its
photons. The results presented here are 
broadly consistent with this general scenario, although there are some specific
inconsistencies or complexities.

A natural consequence of this scenario is that as the
inner radius of the disk moves inwards the relative energy released 
between the inner hot region and the disk decreases. Hence the required
Compton amplification decreases which translates into an increase in the
spectral index $\Gamma$. At the same time, the reflection parameter 
(and consequently the equivalent width of the Iron line) should increase
and hence there should be a positive correlation between these quantities
and $\Gamma$. Furthermore, as the disk moves inwards, the increased relativistic
effects should broaden the Iron line and hence the width of the line
should also be correlated with $\Gamma$. As the disk fills the inner
regions, the reflection and the line width should saturate to
their maximum values and hence in the soft state they should not
be correlated with $\Gamma$. These predictions are broadly consistent with
the results shown in the top and middle panels of Figure \ref{gam_Fe}. Both the 
relative strength (i.e. the Equivalent width) and the width of the Iron line increase
with $\Gamma$ and then saturate. However, the value of $\Gamma$ for which this saturation
occurs is different. Given the spectral resolution of the PCA and the approximate Gaussian
model used for the skewed broad Iron line, one may expect that the Equivalent Width is
a better measured quantity than the width of the line. 
It is interesting to note that
there is a rather sharp discontinuity in the Equivalent Width variation at $\Gamma = 2$.
As the disk extends to the innermost region, the geometry of the
Comptonization region changes from being a hot inner disk with
$\Gamma < 2$ to  a patchy corona on top of the disk with $\Gamma > 2$.
This geometrical transformation at $\Gamma \sim 2$ could be the cause
of the rather sharp discontinuity.
The reflection parameter $R$ has a complex non-monotonic behaviour
with $\Gamma$ (Bottom panel of Figure \ref{gam_Fe}). It is correlated for $\Gamma < 1.8$,
inversely correlated for $1.8 < \Gamma < 2.1$ and correlated for larger values. This
suggests that as the standard disk extends to the innermost radii, the geometry and nature
of the active coronal regions becomes more complex and the disk maybe getting ionised, an
effect which is not taken into account here.

The top panel of Figure \ref{gam_Flux}, shows the expected result that the flux of
the Comptonizing components (filled circles) decreases with increasing $\Gamma$ while
the disk black-body flux increases (open squares). The flux of each of these components is single valued for a
given $\Gamma$ which shows that source {\it never} undergoes any hysteresis effect, confirming results obtained
from a single transition\citep{Zdz04}.   Such an effect has been observed
in some black hole systems (e.g. GX 339-4) where the variation of $\Gamma$ with  flux depends on whether
the source is transiting from low to hard or from hard to low \citep{Zdz04}.
It is interesting to note that for
the entire soft state ($\Gamma > 2.0$) the disk emission is nearly equal to the
Comptonizing one which implies that the fraction of energy dissipated in the corona
is $\sim 0.5$ independent of $\Gamma$. This is consistent with
the results obtained from analysis of a single soft state data in 1996 \citep{Gie99}.
Here we demonstrate the universality of this result for different epochs of the soft state.
It is also clear the Cyg X-1 never exhibits a ultra-soft state spectrum where the disk emission
dominates the Comptonized one. The input flux of soft photons entering the Comptonized region
are marked as open triangles in the top panel of Figure \ref{gam_Flux}. For $\Gamma > 2.0$, this flux
is less than the disk emission, which suggests that the in the soft state the corona is a patchy
one, covering a  fraction of the surface area of the disk. These patchy active regions Comptonize
only a fraction of the disk photons while the rest are observed directly. For the hard state
($\Gamma < 1.8$), it is surprising to note that the input photons have significantly more flux than
the disk emission.   In the truncated disk geometry, one would expect a similar flux of disk photons
entering into the inner Comptonizing region as what is observed directly. This may indicate that the
truncated disk perhaps penetrates the inner hot region.  The temperature of the inner disk,
$kT_{in}$ (bottom panel of Figure \ref{gam_Flux}) also shows an unexpected behaviour. In the soft
state the temperature should be roughly constant, but as the disk recedes in the hard state, 
$kT_{in}$ is expected to decrease. An opposite trend is observed with $kT_{in}$ increasing with 
decreasing $\Gamma$, albeit with a large scatter. Although, such high input disk temperatures for the hard 
state have been reported by \cite{Now11}, the presence of a soft excess, which may be due to
an additional Comptonization component \citep{Ibr05} make interpreting the results difficult.
Another possibility is that the temperature of the inner disk and the seed photon temperature which are
assumed to be same in the model fitting may not be so.
Moreover, it should 
be noted that these results are based on energies $> 3$ keV and on simple empirical 
spectral fitting models and hence need to be confirmed by broadband detailed spectral fitting. 

The amplification factor shows the expected trend of increasing sharply
with decreasing $\Gamma$ in the hard state (top panel of Figure \ref{gamA}). It is
interesting to note that this increase is slightly sharper than the expected curves when $kT_{in}$ is
held constant (solid lines). As the truncated disk recedes the relative energy dissipation between 
the inner region and the disk increases, which requires a larger amplification factor, $A$ for energy 
balance. However, as can be inferred from the figure, the larger $A$ requirement 
does not necessarily correspond to a significant change in $\Gamma$. For a range of $A > 7$, the
spectral index, $\Gamma \sim 1.65$. This naturally explains the absence of spectra with $\Gamma < 1.6$
seen in the distribution of $\Gamma$ (bottom panel of Figure \ref{gamA}).

In summary, the comprehensive spectral analysis of Cyg X-1 using 504 pointed observations, is broadly
consistent with the picture that in the hard state there is a truncated disk and a hot inner region.
As the system moves to the soft state, the inner radius of the disk moves inwards. Finally in the
soft state the hot inner region disappears, with the appearance of active coronal regions on top
of the cold disk. This scenario is supported by the increase of the relative strength and
width of the Iron line with spectral slope $\Gamma$ for small values of $\Gamma < 2$ 
of the hard state 
and their saturation for  large $\Gamma > 2$ values of the soft state. However, the reflection
parameter $R$ shows non-monotonic behaviour with $\Gamma$, indicating complexities in the geometry
or in the ionisation state of the reflector. The analysis confirms that in the soft state the
disk flux is similar to the Comptonized component for a large range of $2 < \Gamma < 2.9$ and
the source never goes into a ultra-soft state. The input photon flux in the soft state is significantly
 smaller than the disk emission, indicating a patchy corona geometry. The inner disk temperature 
remains nearly constant at the soft state but unexpectedly shows an increase in the hard state,
albeit with large scatter.  The Compton Amplification factor increases steeply with 
$\Gamma$ providing a natural explanation for the absence of any spectra with $\Gamma < 1.6$. Finally,
no hysteresis effect is detected, with the spectral index $\Gamma$ being uniquely correlated with 
other parameters including the flux.  The inconsistencies or complexities from the standard paradigm,  indicated in this work need to be
confirmed 
by broad band sensitive instruments especially those covering the spectrum at lower energies 
$E < 3$ keV. Correlated temporal analysis with theoretical understanding of the variability 
may also provide valuable insight in the nature of the system. Overall, with it relative 
simplicity compared to other black hole systems, especially transient ones, 
Cyg X-1 remains a promising source to understand the nature of accreting black hole systems.

\section{Acknowledgements} 

This work has been partially funded by the ISRO-RESPOND program No:ISRO/RES/2/353/08-09. 
SR and SNAJ acknowledge the
IUCAA associateship and visitor's program.

{}


\end{document}